\documentclass[conference]{IEEEtran}
\IEEEoverridecommandlockouts
\usepackage{cite}
\usepackage{amsmath,amssymb,amsfonts}
\usepackage{textcomp}
\usepackage{xcolor}

\usepackage{booktabs} % For formal tables
\usepackage{indentfirst} %two space at the begin of paragragh
\usepackage{graphicx}  %insert graft
\usepackage{makecell}
\usepackage{graphics}
\usepackage{subfigure}
\usepackage{algorithm}
\usepackage{algpseudocode}    
\def\BibTeX{{\rm B\kern-.05em{\sc i\kern-.025em b}\kern-.08em
    T\kern-.1667em\lower.7ex\hbox{E}\kern-.125emX}}
\begin{document}

\title{Adaptive Wavelet Clustering for Highly Noisy Data\\
%\thanks{This work is supported by National Natural Science Foundation (61772219). }
\thanks{This work is supported by National Natural Science Foundation (61772219,61572221). }
}

\author{
	\IEEEauthorblockN{Zengjian Chen}
	%\authornote{The three authors contributing equally.}
	\IEEEauthorblockA{\textit{Department of Computer Science} \\
		\textit{Huazhong University of }\\
		\textit{Science and Technology}\\
		Wuhan, China \\
		m201873061@hust.edu.cn}
	\and
	\IEEEauthorblockN{Jiayi Liu}
	\IEEEauthorblockA{\textit{Department of Computer Science} \\
		\textit{University of Massachusetts Amherst}\\
		Massachusetts, USA \\
		liu@umass.edu}
	\and
	\IEEEauthorblockN{Yihe Deng}
	\IEEEauthorblockA{\textit{Department of Mathematics} \\
		\textit{University of California, Los Angeles}\\
		California, USA \\
		yihedeng@g.ucla.edu}
	\and
	\IEEEauthorblockN{~~~~~~~~}
	\IEEEauthorblockA{~~~~~~~~}
	\and
	\IEEEauthorblockN{ Kun He\textsuperscript{*} 
		\thanks{*: Corresponding author.}
	}
	\IEEEauthorblockA{\textit{Department of Computer Science} \\
		\textit{Huazhong University of Science and Technology}\\
		Wuhan, China \\
		brooklet60@hust.edu.cn}
	\and
	\IEEEauthorblockN{John E. Hopcroft}
	\IEEEauthorblockA{\textit{Department of Computer Science} \\
		\textit{Cornell University}\\
		Ithaca, NY, USA \\
		jeh@cs.cornell.edu}
}

\maketitle

	\vspace{-0.5em}	
\begin{abstract}
In this paper we make progress on the unsupervised task of mining arbitrarily shaped clusters in highly noisy datasets, which is a task present in many real-world applications. Based on the fundamental work that first applies a wavelet transform to data clustering, we propose an adaptive clustering algorithm, denoted as AdaWave, which exhibits favorable characteristics for clustering. By a self-adaptive thresholding technique, AdaWave is parameter free and can handle data in various situations. It is deterministic, fast in linear time, order-insensitive, shape-insensitive, robust to highly noisy data, and requires no pre-knowledge on data models. Moreover, AdaWave inherits the ability from the wavelet transform to cluster data in different resolutions. We adopt the ``grid labeling'' data structure to drastically reduce the memory consumption of the wavelet transform so that AdaWave can be used for relatively high dimensional data. Experiments on synthetic as well as natural datasets demonstrate the effectiveness and efficiency of our proposed method.
\end{abstract}

\begin{IEEEkeywords}
Clustering, high noise data, wavelet transform, shape-insensitive
\end{IEEEkeywords}

\section{Introduction}
\label{sec:Intro}
Clustering is a fundamental data mining task that finds groups of similar objects while keeping dissimilar objects separated in different groups or in the group of noise (noisy points)~\cite{jain1999survey, Kriegel2011Survey}. 
The objects can be spatial data points, feature vectors, or patterns. Typical clustering techniques include centroid-based clustering~\cite{jain2010data}, spectral clustering~\cite{von2007tutorial}, density based clustering~\cite{Kriegel2011Survey}, etc. These techniques usually perform well on ``clean'' data. However, they face a big challenge when dealing with real-world applications where patterns are usually mixed with noise. %``swim'' in a sea of clutter or noise. 
Furthermore, complex and irregualarly shaped groups render these once-effective clustering techniques intractable, because typical clustering approaches are either deficient of a clear ``noise'' concept or limited in specific situations due to its shape-sensitive property.

This work addresses the problem of effectively uncovering arbitrarily shaped clusters when the noise is extremely high. Based on the pioneering work of Sheikholeslami that applies wavelet transform, originally used for signal processing, on spatial data clustering~\cite{sheikholeslami1998wavecluster}, we propose a new wavelet based algorithm called AdaWave that can adaptively and effectively uncover clusters in highly noisy data. 
To tackle general applications, we assume that the clusters in a dataset do not follow any specific distribution and can be arbitrarily shaped.

\begin{figure}[htbp]
\centering
\subfigure[raw data]{
\begin{minipage}{4cm}
\centering
\includegraphics[scale=0.1]{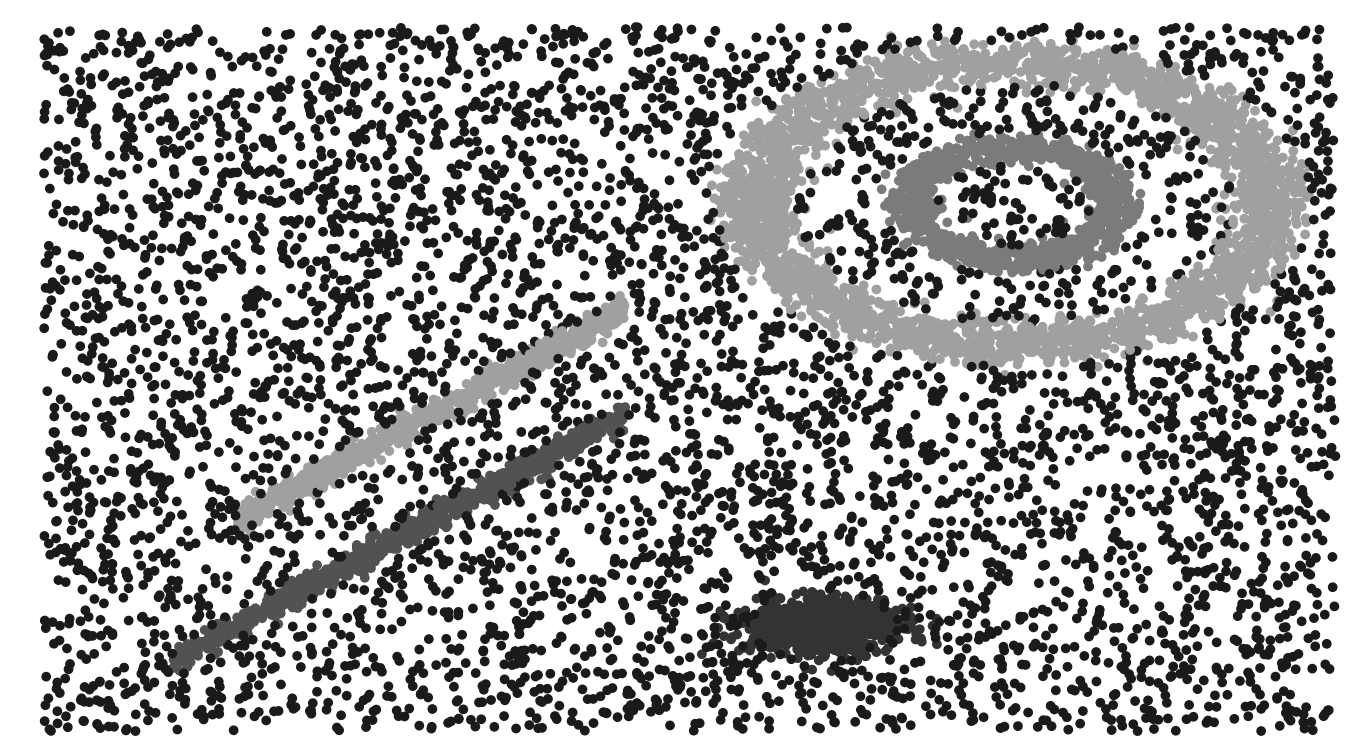}
\end{minipage}
}
\subfigure[AdaWave clustering]{
\begin{minipage}{4cm}
\centering
\includegraphics[scale=0.1]{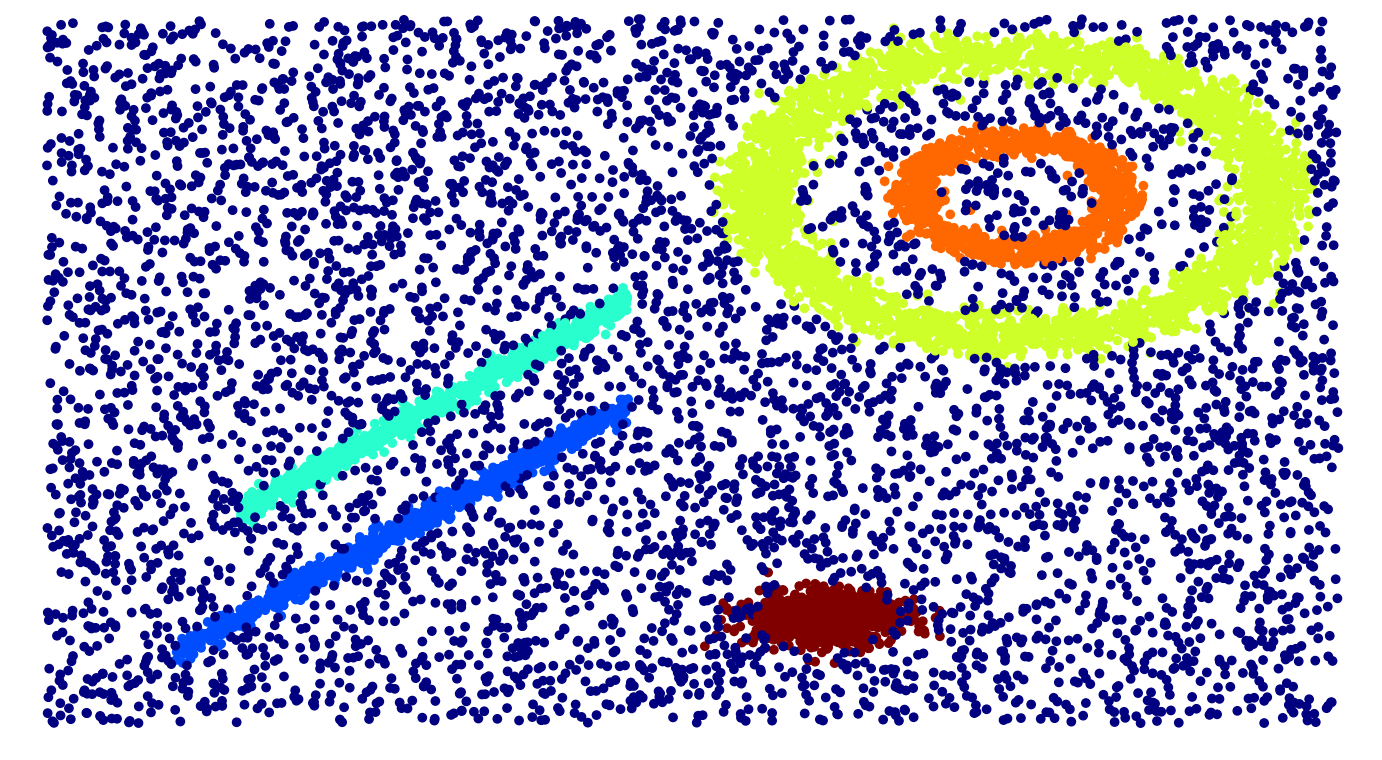}
\end{minipage}
}
\vspace{-1em}
\caption{A running example.}
\vspace{-0.5em}
\label{Fig:AdaWaveExample}
\end{figure}

To show the hardness of the clustering task, we first design a highly noisy running example with clusters in various types, as illustrated in Fig. \ref{Fig:AdaWaveExample}(a).
Fig. \ref{Fig:AdaWaveExample}(b) shows that AdaWave can correctly locate these clusters. 
Without any estimation of the explicit models on the data, the proposed AdaWave algorithm exhibits properties that are favorable to large scale real-world applications. 

\vspace{-0.4em}
 
For comparison, we illustrate the results of several typical clustering methods on the running example: $k$-means~\cite{Steinhaus1957kmeans,Forgy1965kmeans} as the representative for centroid-based clustering methods, DBSCAN~\cite{DBSCAN1996KDD} as the representative for density-based clustering methods~\cite{Kriegel2011Survey,Ankerst99OPTICS}, and SkinnyDip~\cite{maurus2016skinny} as a newly proposed method that handles extremely noisy environments. Besides illustration, we also use Adjusted Mutual Information (AMI)~\cite{bohm2010clustering}, a standard metric ranging from 0 at worst to 1 at best, to evaluate the performances.

\begin{figure*}[ht]
\vspace{-0.5em}
\centering
\includegraphics[scale=0.35]{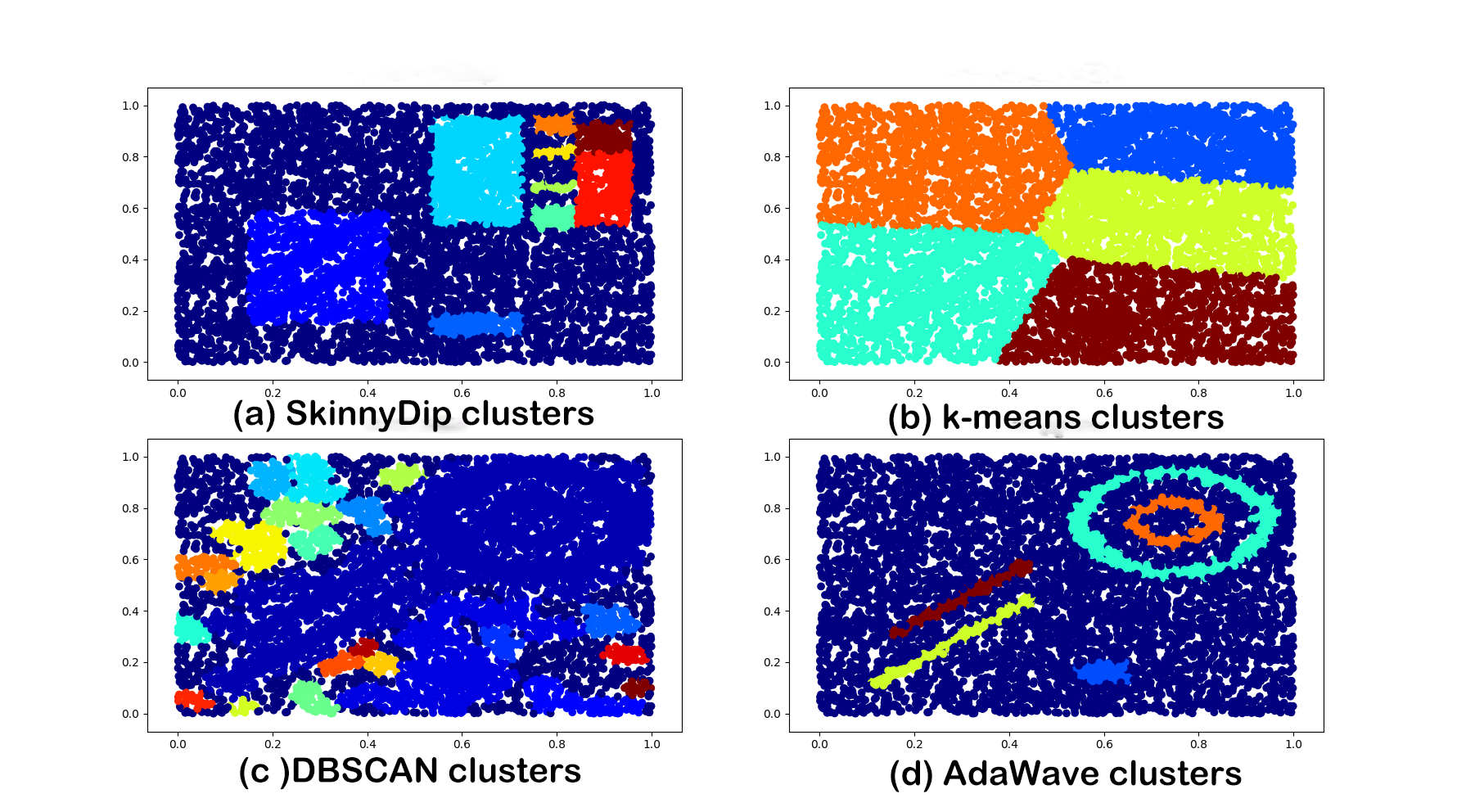}
\vspace{-1em}
\caption{Clutering results on the running example.}
\label{fig:exampleResults}
\vspace{-0.5em}
\end{figure*}

%\vspace{-0.5em}
\begin{itemize} 
	\setlength{\itemsep}{0pt}
	\setlength{\parsep}{0pt}
	\setlength{\parskip}{0pt}	
	\item Centroid-based clustering algorithms tend to lack a clear notion of ``noise'', and behave poorly on very noisy data. As shown in Fig.  \ref{fig:exampleResults}(b), the standard $k$-means yields poor results and the AMI is very low at 0.25.  
	\item Density-based clustering algorithms usually perform well in the case where clusters have heterogeneous form. As a representative, DBSCAN is known to robustly discover clusters of varying shapes in noisy environments. After fine-tuning the parameters that requires graphical inspection, the result of DBSCAN is shown in Fig. \ref{fig:exampleResults}(c). DBSCAN roughly finds the shapes of the clusters, but there is still much room for improvement. It detects 21 clusters with an AMI of 0.28, primarily because there are various areas in which, through the randomness of noise, the local density threshold is exceeded. Also, we have tested various parameters and did not find a parameter combination for DBSCAN that can solve the problem. 
	\item SkinnyDip is known to robustly cluster spatial data in extremely noisy environments. For the running example, we sample from the nodes of a sparse grid \cite{bungartz2004sparse}, which is regarded as the best mechanism for choosing a starting point for gradient-ascent in SkinnyDip. However, the clustering result in Fig. \ref{fig:exampleResults}(c) shows that Skinnydip performs poorly as the datasets do not satisfy its assumption that the projection of clusters to each dimension is in a unimodal shape. 
\end{itemize}
 
The above methods heavily rely on the estimation of explicit models on the data, and produce high quality results when the data is organized according to the assumed models. However, when facing more complicated and unknown cluster shapes or highly noisy environments, these methods do not perform well as expected. 
By comparison, as shown in Fig. \ref{fig:exampleResults}(d), the proposed AdaWave, which applies wavelet decomposition to denoising and ``elbow theory'' in adaptive threshold setting, correctly detects all the five clusters and groups all the noisy points as the sixth cluster. AdaWave achieves an AMI value as high as 0.76, and furthermore, computes deterministically and runs in linear time.

We propose AdaWave\footnote{The data and code will be public after the review.} to cluster highly noisy data which frequently appears in real-world applications where clusters are in extreme-noise settings with arbitrary shapes. We also demonstrate the efficiency and effectiveness of AdaWave on synthetic as well as real-world data. 
The main characteristics of AdaWave that distinguish the algorithm from existing clustering methods are as follows: 

\vspace{-0.5em}
\begin{itemize} 
	\setlength{\itemsep}{0pt}
	\setlength{\parsep}{0pt}
	\setlength{\parskip}{0pt}	
	\item AdaWave is deterministic, fast, order-insensitive, shape-insensitive, robusts in highly noisy environment, and requires no pre-knowledge on the data models. To our knowledge, there is no other clustering method that meets all these properties.
	\item We design a new data structure for wavelet transform such that comparing to the classic WaveCluster algorithm, AdaWave is able to handle high-dimensional data meanwhile remain storage-friendly in such situation. 
	\item We propose a heuristic method based on \textit{elbow theory} to adaptively estimate the best threshold for noise filtering. By implementing the self-adaptive operation on the threshold, AdaWave exhibits high robustness in extremely noisy environment, and outperforms the state-of-art baselines by experiments.
\end{itemize}

%The rest of this paper is organized as follows.
% Section \uppercase\expandafter{\romannumeral2} provides a quick review for existing clustering methods.
% Section \uppercase\expandafter{\romannumeral3} analyzes the properties of wavelet transform and explains why wavelet transform exhibits distinctive strength in clustering and automatical denoising. In Section \uppercase\expandafter{\romannumeral4} we discuss the $elbow$ $theory$ in detail and summarize the proposed clustering algorithm AdaWave. Section \uppercase\expandafter{\romannumeral5} carries out experiments and does comparisons with related clustering methods. The results are discussed in Section \uppercase\expandafter{\romannumeral6}, Section \uppercase\expandafter{\romannumeral7} concludes our work. 

\section{Related Work}
\label{sec:Relate}
Various approaches have been proposed to improve the robustness of clustering algorithms in noisy data~\cite{DBSCAN1996KDD, dasgupta1998detecting,maurus2016skinny}.
Here we highlight several algorithms most related to this problem and focus on illustrating the preconditions for these clustering methods.

DBSCAN~\cite{DBSCAN1996KDD} is a typical density-based clustering method designed to reveal clusters of arbitrary shapes in noisy data. When varying the noise level on the running example, we find that DBSCAN performs well only when the noise is controlled below 15\%. Its performance derogates drastically as we continue to increase the noise percentage. Meanwhile, the overall average time complexity for DBSCAN is $O(N \text{log} N)$ for $N$ data points and, in the worst case, $O(N^2)$. Thus, its running time can also be a limiting factor when dealing with large scale datasets. Another density-based clustering method, Sync~\cite{bohm2010clustering}, exhibits the same weakness on time complexity (reliance on pair-wise distance, with time complexity of $O(N^2)$.

Regarding data with high noise, as early as 1998, Dasgupta et al.~\cite{dasgupta1998detecting} proposed an algorithm to detect minefields and seismic faults from ``cluttered'' data. Their method is limited to the two-dimensional case, and an extended version for slightly higher dimension ($d \leq 5$) requires significant parameter tuning.

In 2016, Samuel et al.~\cite{maurus2016skinny} proposed an intriguing method called SkinnyDip~\cite{maurus2016skinny}. SkinnyDip optimizes DipMeans~\cite{kalogeratos2012dip} with an elegant dip test of unimodality~\cite{hartigan1985dip}. It focuses on the high noise situation, and yields good result when taking a unimodal form on each coordinate direction. However, the condition is very strict that the projections of clusters have to be of unimodal shapes in every dimension. When such condition does not exist, SkinnyDip could not uncover clusters correctly. 

A newly proposed work in 2017~\cite{Bojchevski2017KDD} applies a sparse and latent decomposition of the similarity graph used in spectral clustering to jointly learn the spectral embedding and the corrupted data. It proposes a robust spectral clustering (RSC) technique that are not sensitive to noisy data. Their experiments demonstrated the robustness of RSC against spectral clustering methods. However, it can only deal with low-noise data (up to $10\%$ and $20\%$ of noise).

Our proposed AdaWave algorithm targets extreme-noise settings as SkinnyDip does.
Besides, as a grid-based algorithm, AdaWave shares the common characteristic with STING~\cite{wang1997sting} and CLIQUE~\cite{Agrawal98sigmod}: fast and independent of the number of data objects. Though the time complexity of AdaWave is $O(nm)$, where $m$ is the number of grids. The complexity is slightly higher than that of SkinnyDip, but AdaWave still runs in linear time and yields good results when dataset consists of clusters in irregular shapes.

Experiments in Section \uppercase\expandafter{\romannumeral5} show that AdaWave outperforms other algorithms, especially in the following scenarios that happen commonly in large scale real applications, when the data 1) contains clusters of irregular shapes such as rings, 2) is a very large dataset in relatively high dimensions, 3) contains a very high percentage (for example 80\%) of noise.

\section{Wavelet Transform}
\label{sec:Wavelet}
Wavelet transform has been known as an efficient denoising technology. Overpassing its predecessor Fourier Transform, wavelet transform can analyze the frequency attributes of a signal when its spatial information is retained.

\subsection{Principles of Wavelet Transform}

In this section, we focus on $Discrete$ $Wavelet$ $Transform$ (DWT) which is applied in AdaWave algorithm. The `transform' in DWT separates the original signal into scale space and wavelet space. The scale space stores an approximation of the outline of the signal while the wavelet space stores the detail of the signal. Referring to the Mallat algorithm \cite{Mallat89Pami}, we can simplify the complicated process of DWT into two filters.

\subsubsection{1D Discrete Wavelet Transform}
\begin{figure}[htbp]
	\centering
\vspace{-0.5em}	
	\includegraphics[scale=0.6]{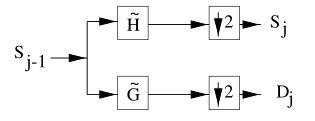}
	\includegraphics[scale=0.6]{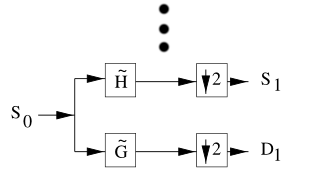}
\vspace{-0.5em}	
	\caption{Mallat Algorithm~\cite{sheikholeslami1998wavecluster}. }
\vspace{-0.5em}	
	\label{fig:dwt_filter}
\end{figure}

As shown in Fig. \ref{fig:dwt_filter}, signal $S_{j-1}$ pass two filters $\widetilde{H}$ and $\widetilde{G}$ and is down-sampled by 2. According to the Mallat algorithm\cite{sheikholeslami1998wavecluster}, signal can be decomposed into scale space and wavelet space by passing a low-pass filter and a high-pass filter correspondingly. Choosing different wavelet functions, we can get related filters by looking up a precalculated table. 

\begin{figure}[htbp]
	\centering
	\vspace{-0.5em}	
	\includegraphics[scale=0.35]{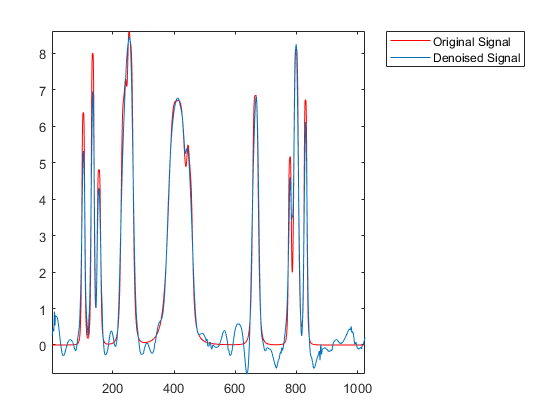}
	\vspace{-1em}
	\caption{Illustration on denoising.}
	\label{fig:denoise}
	\vspace{-1em}	
\end{figure}

Just by passing the signal through two filters and down sample by 2, we are able to decompose the signal into a space that only contains the outline of the signal and another space that only contains the detail. The signal discussed here includes high dimensional signals. Passing a filter for a $d$-dimensional signal is just repeating the process of 1D signal for $d$ times.

In both data science and information theory, noise is defined as the collection of unstable, nonsense points in signal (dataset). As shown in Fig.  \ref{fig:denoise}, in denoising task, we are trying to maintain the outline of the signal and amplify the contrast between high value and low value, which is a perfect stage for DWT.

\subsubsection{2D Discrete Wavelet Transform}

\begin{figure}[htbp]
	\centering
	\subfigure[original feature space]{
		\begin{minipage}{4cm}
			\centering
			\includegraphics[scale=0.6]{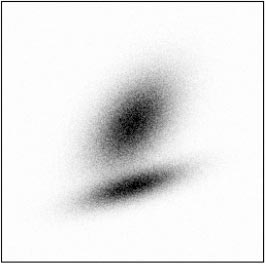}
		\end{minipage}
	}
	\subfigure[transformed feature space]{
		\begin{minipage}{4cm}
			\centering
			\includegraphics[scale=0.6]{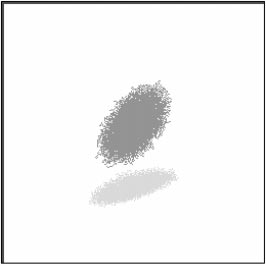}
		\end{minipage}
	}
	\vspace{-0.8em}
	\caption{2D discrete wavelet transform.}
	\vspace{-0.5em}
	\label{fig:twodDWT}
\end{figure}

To further show DWT's denoising feature regarding space data, we apply two separate 1D wavelet transform on 2D dataset and filter co-efficient in transformed feature space. Referring to~\cite{Michael1994Multimedia}, the 2D feature space is first convolved along the horizontal ($x$) dimension, resulting in a low-pass feature space $L$ and a high-pass feature space $H$. We then downsample each of the convolved space in the $x$ dimension by 2. Both $L$ and $H$ are then convolved along the vertical ($y$) dimension, resulting in four subspaces: $L_{x}L_{y}$ (average signal), $L_{x}H_{y}$ (horizontal features), $H_{x}L_{y}$ (vertical features), and $H_{x}H_{y}$ (diagonal features). Next, by extracting the signal part ($L_{x}L_{y}$) and discarding low-value coefficients, we obtain the transformed feature space, as shown in Fig. \ref{fig:twodDWT}.

Intuitively, according to Fig. \ref{fig:twodDWT}, dense regions in the original feature space act as attractors to the nearby points and at the same time as inhibitors to the points that are not close enough. This means that clusters in the data and the clear regions around them automatically stand out and become more distinct. Also, it is evident that the number of points sparsely scattered (outliers) in the transformed feature space is lower than that in the original space. The decrease in outliers reveals the robustness of DWT regarding extreme noise.

\vspace{-0.5em}

As mentioned earlier, the above wavelet model can similarly be generalized in $d$-dimensional feature space, where one-dimensional wavelet transform will be applied $d$ times.

\subsection{Properties of Wavelet Transform}

Besides its great ability of analyzing frequency and spatial information at the same time, other properties of DWT also distinguish it among various denoising methods.

$\bullet$ \textbf{Low entropy.} The sparse distribution of wavelet coefficients reduces the entropy after DWT. Thus after the signal decomposition, many wavelet coefficients are close to zero, which generally refers to the noise. The main component of the signal is more concentrated in some wavelet basis, therefore removing the low-value coefficients is an effective denoising method that can better retain the original signal.

$\bullet$ \textbf{Multi-resolution.} As shown in Fig. \ref{fig:dwt_filter}, DWT is implemented in a layered structure. In each layer of decomposition, we only decompose $S_j$ which represents the wavelet space, also known as the detailed feature of the signal, and preserve the scale space $D_j$. Such layered structure gives us the possibility to observe signal in different resolutions of scale space, $D_0$ to $D_j$, in a single application of DWT.

$\bullet$ \textbf{De-correlation.} As illustrated above, DWT can separate signals into scale space and wavelet space. By such separation, DWT de-correlated the `detail' (the part that oscillated very fast in Fig. \ref{fig:denoise}). With the de-correlation property, DWT works especially well on separating noise from highly noisy data.

$\bullet$ \textbf{Flexibility of choosing basis.} Wavelet transform can be applied with a flexible choice of wavelet function. As described above, for each kind of wavelet function, there exists a pre-calculated filter. According to different requirements, various wavelet families have been designed, including Daubechies, Biorthogonal and so on. The users of AdaWave have the flexibility to choose any kind of wavelet basis, which makes the algorithm universal under various requirements.

\section{Wavelet Based Adaptive Clustering}
\label{sec:AdaWave}
Given a dataset with $n$ data points in $d$ dimensional space, and the coordinate of each data $a_i$ is ($a_{i1},...,a_{id}$), labeled in the ground-truth as  either in a cluster or noise, the clustering problem is to cluster the data and filter the noise. 
AdaWave is an efficient algorithm based on wavelet transform. It is generally divided into four parts, as shown in Algorithm 1. 

At the first step we propose a new data structure for clustering high-dimensional data. With space quantization, we divide the data into grids and project them to a high-scale space. Step 2 is a wavelet transform that preliminarily denoises by removing wavelet coefficients close to zero. Step 3 is threshold filtering, which is a further step to eliminate noise, and we apply ``elbow theory'' in adaptively setting the threshold. At the last step, we label and make the lookup table, thereby transforming grids to the original data.  

\begin{algorithm}[h]  
	\caption{AdaWave algorithm}  
	\begin{algorithmic}[1] 
		\Require  
		Data Matrix    
		\Ensure  
		Clustered objects
		\State Quantize feature space, then assign objects to the grids.
		\State Apply wavelet transform to the quantized feature space.
		\State Adaptively find the threshold and filter the noise.
		\State Find the connected components (clusters) in the subbands of transformed feature space at different levels.
		\State Assign labels to the grids.
		\State Make the lookup table and map objects to clusters.    
		\label{code:overall_algo}  
	\end{algorithmic}  
\end{algorithm}

\subsection{Quantization}

The first step of AdaWave is to quantize the feature space. Assume that the domain $B_{j}$ at the $j^{th}$ dimension in the feature space is divided into $m_i$ intervals. By making such division in each dimension, we separate the feature space into multiple grids. Objects (data points) are allocated into these grids according to the coordinates at each dimension. 

In the original wavelet clustering algorithm WaveCluster\cite{sheikholeslami1998wavecluster}, each grid $g_{i}$ is the intersection of one interval from each dimension and $g_{ij}=[l_{ij},h_{ij})$ is a right open interval in the partition of $B_{j}$. An object (data point) \textbf{$a_k$}=$ \langle a_{k1},a_{k2},...,a_{kd} \rangle $ is contained in a grid \textbf{$g_{i}$} if $l_{ij}$ $\leq$ $a_{kj} < h_{ij}$ for 1 $\leq$ $j$ $\leq$ $d.$
For each grid, we use the number of objects contained in the grid as the value of grid density to represent its statistical feature. The selection of the number of grids to generate and statistical feature to use can significantly affect the performance of AdaWave.

It is easy to get trapped by storing all grids in the feature space to keep the quantized data. Even though quantifying the data can be completed in linear time, it can lead to an exponential memory consumption with regard to dimension $d$. In AdaWave, we successfully achieve the goal of ``only storing the grid with non-zero density'' by ``labeling'' the grids in the data space. When considering low dimensional data, only storing grids with non-zero density cannot demonstrate its advantage because of the high data density in the entire space. However, in high dimension, the number of grids far exceeds the number of data points. When a lot of grids are of zero density, the above strategy can save considerable amount of memory, making it possible to apply AdaWave to high dimensional clustering problems.

\begin{algorithm}[h]  
	\caption{Data space quantification}  
	\begin{algorithmic}[1] 
		\Require  
		Data Matrix $A \in \mathbb{R}^{n\times d}$, each row $a_i$ corresponds to the coordinates of object $i$
		\Ensure  
		Quantized grid set $G \in \mathbb{R}^{d\times d}$, stored as \{id:density\}
		\State $G = \O$
		\For {$a_i$ ($i$ $\in$ \{1,...$n$\})}
		\State /*Calculate the id of the grid*/  
		\State $g^{id}$ = getGridID($a_i$)  
		\If {$g^{id}$ $\in$ $G$} 
		\State /*if $g$ exists in $G$, add 1 to its density*/
		\State $G$.get($g^{id}$) += 1
		\Else
		\State /*else, set its density to 1*/
		\State $G$.add($g^{id}$ = 1)
		\EndIf
		\EndFor  
		\label{code:recentEnd}  
	\end{algorithmic}  
\end{algorithm}

\subsection{Transformation and Coefficient Denoising}

At the second step, %$d$-dimensional discrete wavelet transform will be applied to the quantized feature space. 
we apply wavelet transform to the $d$-dimensional grid set $G = \{g_1,...g_d\}$ to transform the grids into a new feature space. According to Eq. (1), the original data can be represented by wavelet coefficients $c_{j_0,k}$ and scaling coefficients $d_{j,k}$ in the feature space, determined by orthonormal basis $\varphi_{j_0,k}(m)$ and $\psi_{j,k}(m)$.

\begin{equation}
g(m) = \sum_{k}c_{j_0,k}\varphi_{j_0,k}(m) + \sum_{j>j_0}^{}\sum_{k}d_{j,k}\psi_{j,k}(m)
\end{equation}

We calculate the coefficients by Eq. (2), where scaling coefficients \{$d_{j,k}$\} represent the signal after low-pass filtering that preserves the smooth shape, while wavelet coefficients represent the signal after high-pass filtering and usually correspond to the noise part. 
\begin{equation}
\left\{
\begin{array}{lr}
c_{j_0,k}=\langle g(m),\varphi_{j_0,k}(m) \rangle &  \\
d_{j,k} =\langle g(m),\psi_{j_0,k}(m) \rangle &  
\end{array}
\right.
\end{equation}

After discrete wavelet transform, we remove the wavelet coefficients and the low value of scaling coefficients (noise part of the signal), then reconstruct the new grid set $\hat{G}$ with the remaining coefficients. In this way, noise (outliers) is automatically eliminated. %AdaWave detects the connected components in the transformed feature space. Each connected component is a set of grids in \{\textbf{$t_k$} : 1 $\leq$ $k$ $\leq$ $\kappa$\} and is considered as a cluster. Corresponding to each resolution of the wavelet transform, there is a set of clusters, where, usually at the coarser resolutions, the number of clusters is low. 
%In experiments, we applied each of the three-level wavelet transforms, Daubechies, Cohen-Daubechies-Feauveau ((4,2) and (2,2)). $Average$ $subbands$ (feature spaces) give approximations of the original feature space at different scales to find clusters at different levels.

\begin{algorithm}[h]  
	\caption{Wavelet decomposition}  
	\begin{algorithmic}[1] 
		\Require  
		Grids after quantization $G \in \mathbb{R}^{d\times d}$, stored as \{id:density\}
		\Ensure  
		The result after wavelet decomposition $\hat{G} \in \mathbb{R}^{d\times d}$, stored as \{id:density\}
		\State $\hat{G} = \O$
		\State /*wavelet decomposition in each dimension*/ 
		\For{$i$ in \{1,...,$d$\}}
		\State /*calculate the new grid set $\hat{G}_i$ in this dimension*/ 
		\For{each grid $g_i$ $\in$ $G$}  
		\State $\hat{g_i}$ = DWT($g_i$.id, $g_i$.density)
		%\State /*length of the basis may be greater than 2 and cause an overlap.*/
		\If {$\hat{g^i}$ $\in$ $\hat{G_i}$} 
		\State $\hat{G_i}$.get($\hat{g_i}$.id) += $\hat{g_i}$.density
		\Else
		\State $\hat{G_i}$.add($\hat{g_i}$)
		\EndIf
		\EndFor 
		\EndFor  
		\label{code:recentEnd}  
	\end{algorithmic}  
\end{algorithm}

\subsection{Threshold Filtering}
At the third step, a key step of AdaWave, we identify noise by removing the noise grids from the grid set. For a high noise percentage, it is hard for the original WaveCluster to eliminate noise by applying wavelet transform to the original feature space and take advantage of the low-pass filters used in the wavelet transform to automatically remove the noise.

After performing wavelet transform and eliminating the low value coefficients, we can preliminarily remove the outliers with the sparse density. In other words, when the noise percentage is below 20\%, wavelet transform gains outstanding performance, and the computation complexity is $O(N)$. The automatic and efficient removal of the noise enables AdaWave to outperform many other clustering algorithms. 

If 50\% of the dataset is noise, many noise grids would also have high density and purely applying wavelet transform cannot distinguish noise from clusters. Therefore, an additional technique is applied to further eliminate the noise grids.

\begin{figure*}[http]
	\centering
	\subfigure[sorted grids]{
		\begin{minipage}{7cm}
			\centering
			\includegraphics[scale=0.16]{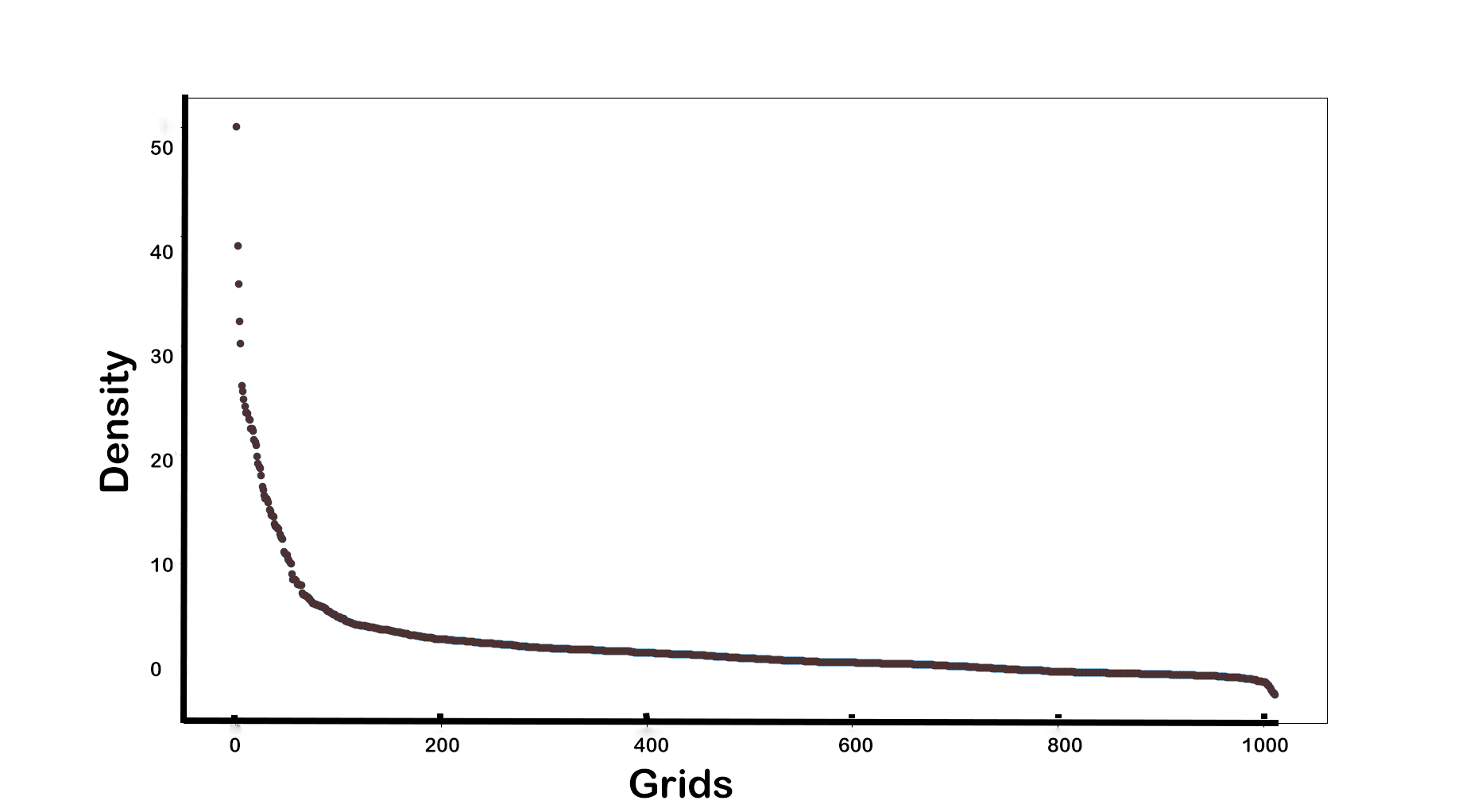}
		\end{minipage}
	}
	\subfigure[adaptively find the threshold]{
		\begin{minipage}{7cm}
			\centering
			\includegraphics[scale=0.16]{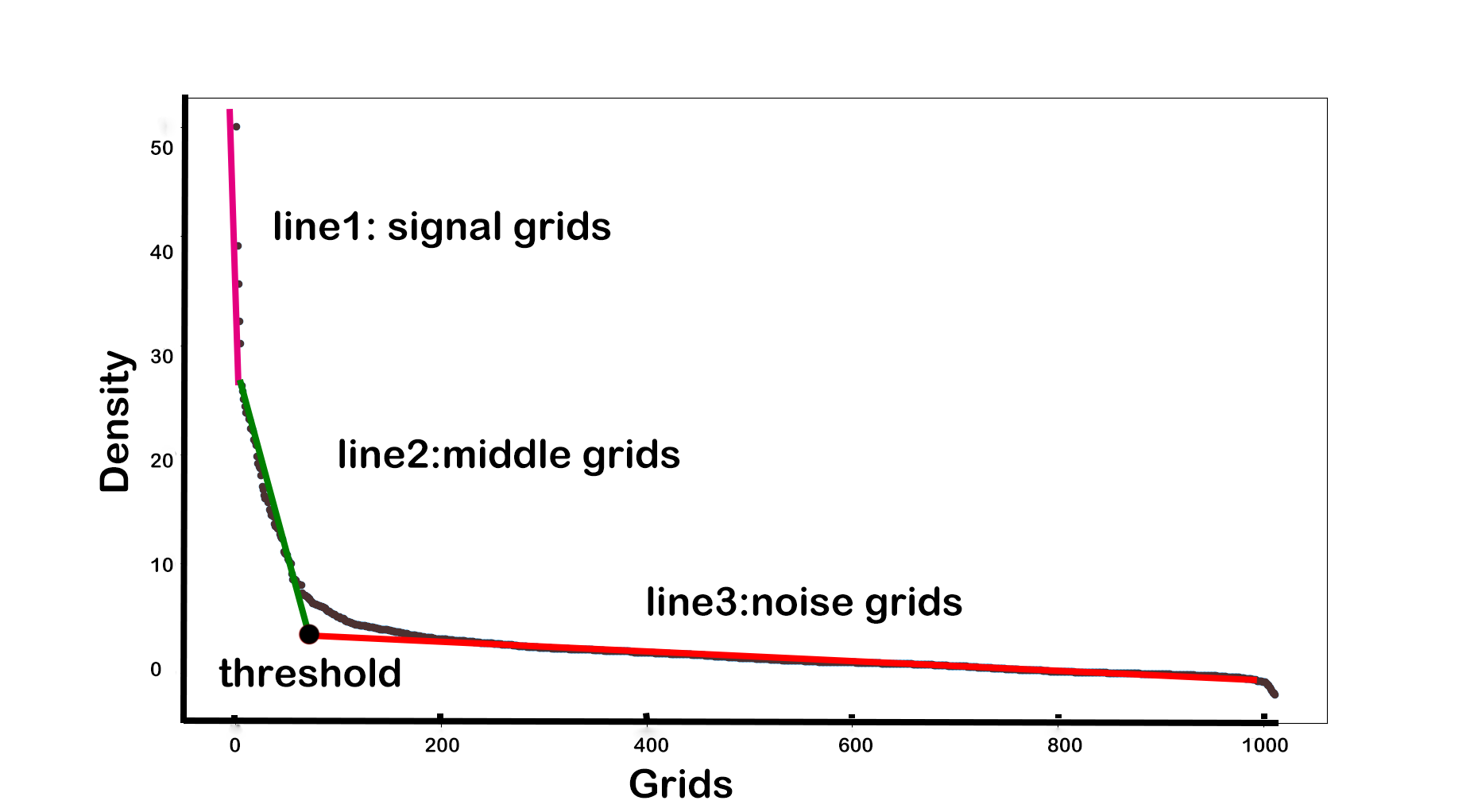}
		\end{minipage}
	}
	\vspace{-0.8em}	
	\caption{Threshold choosing.}
	\vspace{-0.5em}	
	\label{fig:find_thred}
\end{figure*}

The density chart after sorting is shown in Fig. \ref{fig:find_thred}(a) and the grid densities occur as shown because the entire data space is ``averaged'' during the wavelet transform. The essence of the wavelet transform is a process of ``filtering''. Since the filter corresponding to the scale space in wavelet transform is a ``low-pass filter'', the high frequency part that changes rapidly in grid data is filtered out, leaving the low frequency data to represent the signal profile.

After low-pass filtering, the density of the grids in the signal space is roughly divided into three categories: signal data, middle data, and noise data. The chart of these three types of grids is statistically fitted with three line segments. The grid density of signal data should be the largest, represented by the leftmost vertical line in Fig. \ref{fig:find_thred}(b). The middle part consists of the grids between clusters and noise. Due to the low-pass filtering in wavelet transform, these grids has density lower than the grids in the class but higher than that of the noise-only grids. The density of these grids decreases according to its distance from the class. The noise part also appears to be relatively smooth due to the low-pass filtering. Since the density of the noise data is much lower than that of the data in the class, the noise data is represented by the third line that is almost horizontal.

Based on our test experiments on various datasets, the position where the ``middle line'' and the ``noise line'' intersects is generally the best threshold. The algorithm below is the adaptive technique to find the statistically best threshold.

\begin{algorithm}[h]  
	\caption{Thresholding choosing}  
	\begin{algorithmic}[1] 
		\Require  
		Sorted grid set $\widetilde{G}$, stored as \{id:density\}    
		\Ensure  
		The threshold for filtering %$t$
		\State $\theta_{0}$ = $\pi$;  
		\State /*scan the sorted grids*/ 
		\For{$i$ = $\{2,...,|\widetilde{G}-1|\}$}
		\State $\vec{v_1}$ = $\widetilde{G}(i-1)-\widetilde{G}(i)$;  
		\State $\vec{v_2}$ = $\widetilde{G}(i)-\widetilde{G}(i+1)$;
		\State $\theta$ = $arcos\left(\frac{\vec{v_1} \cdot \vec{v_2}}{|\vec{v_1}| |\vec{v_2}|}\right)$
		\State /*renew $\theta_{0}$*/
		\If{$\theta > \theta_{0}$}
		\State $\theta_{0}$ = $\theta$;
		\EndIf
		\State /*find the turning point*/
		\If {$\theta \leq \frac{\theta_{0}}{3}\qquad$} 
		\State	\Return  $\widetilde{G}(i)$.density
		\EndIf
		\EndFor  
		\label{code:recentEnd}  
	\end{algorithmic}  
\end{algorithm} 

\subsection{Label and Make Lookup Table}

Each cluster has a cluster ID. The forth step of the AdaWave labels the grids in the feature space included in a cluster with the cluster ID. 
%That is, $\forall \omega \forall t_k,g_k \in \omega \Rightarrow l_{g_k}=\omega_{id}$, where $l_{g_k}$ is the label of the grid $g_k$. 
The clusters found in the transformed feature space cannot be directly used to define the clusters in the original feature space, because they are only based on wavelet coefficients. AdaWave builds a lookup table $LT$ to map the grids in the transformed feature space to the grids in the original feature space. Each entry in the table specifies the relationship between one grid in the transformed feature space and the corresponding grid(s) in the original feature space. Therefore, the label of each grid in the original feature space can be easily determined. Finally, AdaWave assigns the label of each grid in the feature space to all objects whose feature vector is inside that grid, and thus determines the clusters. 
%Formally, $\forall \omega \forall c_j$ , $\forall o_i \in c_j,l_{o_i} = \omega_{n},\omega \in \zeta_{r}$,$1 \leq i \leq$ $N$, where $l_{o_i}$ is the cluster label of object $O_i$.

\subsection{Time Complexity}
Let $n$ be the number of objects in the database and $m$ be the number of grids. Assuming that the feature vectors of the objects are in $d$-dimensions. Here we suppose $n$ is large and $d$ is comparatively small.

For the first step, the time complexity is $O(nm)$, because AdaWave scans all the objects and assigns them to the corresponding grids, where the domain $B_i$ at each dimension in the $d$-dimensional feature space will be divided into intervals. Assuming the number of intervals $m_i = M$ for each dimension of the feature space, there would be $m = M^d$ grids \cite{wu2006adaptive}.

For the second step, the complexity of applying wavelet transform to the feature space is $O(ldm)= O(dm)$, where  $l$ is a small constant representing the length of the filter used in the wavelet transform. As $d$ is small, we regard $d$ as a constant and  $O(dm)= O(m)$. If we apply wavelet transform to $T$ decomposition levels to sample in each level downward, the required time is less than $O(\frac{4}{3}m)$ \cite{zhao2008research}. That is, the cost of wavelet transform is less than $O(\frac{4}{3}m)$, indicating that a multi-resolution clustering can be very effective. 

The third step aims at finding a suitable threshold for further denoising. The sorting algorithm has a time complexity of $O(mlog(m))$, and filtering takes $O(m)$. Therefore, the total complexity of this step is $O(mlog(m)+m)$ .

To find the connected components in the feature space, the required time will be $O(cm)=O(m)$, where $c$ is a small constant. Making the lookup table requires $O(m)$ time. After reading data objects, data processing is performed sequentially. Thus the time complexity of processing data (without considering I/O) would be $O(m)$, which is independent of the number of data objects $n$. The time complexity of the last step is  $O(m)$. 

For very large datasets, $n \geq m$, $O(n)>O(m)$, the overall time complexity of AdaWave is $O(2*nm + mlog(m) + 2m) = O(nm)$.

\subsection{Properties of AdaWave}
$\bullet$ AdaWave is able to \textbf{detect clusters in irregular shapes}. In AdaWave algorithm, the spatial relationship in data has been preserved during the quantization and wavelet transform. By clustering connected grids into the same cluster, AdaWave makes no assumption on the shape of the clusters. It can find convex, concave, or nested clusters.

$\bullet$ AdaWave is \textbf{noise robust}. Wavelet transform is broadly known for its great performance in denoising. AdaWave takes advantage of this property and can automatically remove the noise and outliers from the feature space. 

$\bullet$ AdaWave is \textbf{memory efficient}. We overcome the problem of exponentially memory grow in wavelet transform for high dimensional data. By using `grid labeling' and the strategy of `only store none-zero grids', AdaWave is able to process data in comparatively high dimensional space, which drastically expands the limit of WaveCluster algorithm.

$\bullet$ AdaWave is \textbf{computationally efficient}. The time complexity of AdaWave is $O(nm)$ where $n$ denotes the number of data points and $m$ denotes the number of grids stored. AdaWave is very efficient for large datasets where $m\ll n$, and $d\ll n$.

$\bullet$ AdaWave is \textbf{input-order insensitive}. When the objects are assigned to the grids in the quantization step, the final content of the grids is independent of the order in which the objects are presented. The following steps of the algorithm will only be performed on these grids. Hence, the algorithm will have the same results with any order of the input data.

$\bullet$ AdaWave can \textbf{cluster in multi-resolution} simultaneously by taking advantage of the multi-resolution attribute from the wavelet transform. By tweaking the quantization parameters or the wavelet basis, users can choose different resolution for clustering.

\section{Experimental Evaluation}
\label{sec:Exp}
In this section, we turn to a practical evaluation of AdaWave. First, according to general classification of acknowledged clustering methods, we choose the state-of-the-art representatives from different families for comparison. Then, we generate a synthetic dataset which exhibits the challenging properties on which we focus and compare AdaWave with the selected algorithms. We apply our method to real-world datasets, and do runtime experiment to evaluate the efficiency of AdaWave.

\subsection{Baselines}

For comparison, we evaluate AdaWave against a set of representative baselines from different clustering families and the state-of-art algorithms. 

We begin with $k$-means~\cite{Steinhaus1957kmeans,Forgy1965kmeans}, which is a widely known technique of centroid-based clustering methods. To achieve the best performance of $k$-means, we set the correct the parameter for $k$. With DBSCAN~\cite{DBSCAN1996KDD}, we have the popular member of the density-based family that is famous for clustering arbitrary shape groups. EM~\cite{CeleuxG1992Elsevier} focuses on probability instead of distance computation; a multivariate Gaussian probability distribution model is used to estimate the probability that a data point belongs to a cluster, with each cluster regarded as a Gaussian model.

Next, we compare AdaWave to advanced clustering methods proposed recently. RIC~\cite{bohm2006robust}, which fine-tunes an initial coarse clustering based on the minimum description length principle (a model-selection strategy based on balancing accuracy with complexity through data-compression). DipMeans~\cite{kalogeratos2012dip} is a wrapper for automating EM and $k$-means respectively. Self-tuning spectral clustering (STSC)~\cite{bohm2010clustering} is a popular automated approach to spectral clustering.  Moreover, we consider SkinnyDip~\cite{maurus2016skinny} because it is a newly proposed method that performs well on handling high-noise. The continuity properties of the dip enable SkinnyDip to exploit multimodal projection pursuit and find an appropriate basis for clustering.

\subsection{Synthetic datasets}

In the synthetic data, we try to mimic the general situation for clustering under very high noise. In the dataset, we simulate different shapes of clusters and various space relations between two clusters. Thus, clusters may not be able to create a uniform shape when projected to a dimension.

\begin{figure}[ht]
\centering
\includegraphics[scale=0.5]{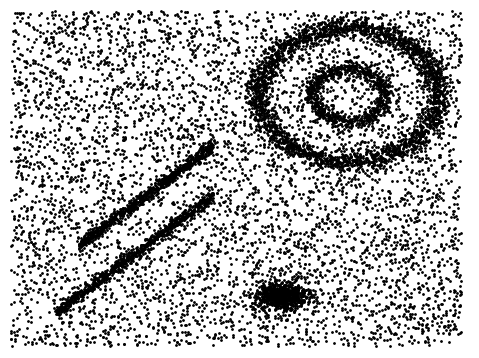}
\caption{A synthetic data set (noise 50\%).}
\label{fig:syn_data}
\end{figure}

By default, we generate five clusters of 5600 objects each in two dimensions, as shown in Fig. \ref{fig:syn_data}. There is a typical cluster roughly within an ellipse with each data in Gaussian distribution with a standard deviation of 0.005. To increase the difficulty of clustering, the next two clusters are of circular distributions overlapping in the directions of $x$ and $y$. The remaining two clusters are in the shape of parallel sloping lines. To evaluate AdaWave with baselines at different degree of noise, we systematically vary the noise percentage $\gamma$ = \{20,25, ... ,90\} by sampling from the uniform distribution over the whole square. In Fig. \ref{fig:syn_data}, the noise is 50\%.

\vspace{-0.5em}	

As a parameter-free algorithm, AdaWave uses its default value of $scale = 128$ with parameters in all cases, and we choose Cohen-Daubechies-Feauveau (2,2) for the wavelet transform. We likewise use the default parameter values for the provided implementations of the baseline algorithms which require neither obscure parameters nor additional processing. To automate DBSCAN, we fix $minPts$= 8 and run DBSCAN for all $\epsilon = \{0.01,0.02,...,0.2\}$, reporting the best AMI result from these parameter combinations in each case. For $k$-means, we similarly set the correct $k$ to achieve automatic clustering and ensure the best AMI result.
% set  $k = \{1,2,...,10\}$ as the cluster number to explore the best AMI result.

%%%%%%%

\begin{figure}[ht]
\centering
\includegraphics[scale=0.35]{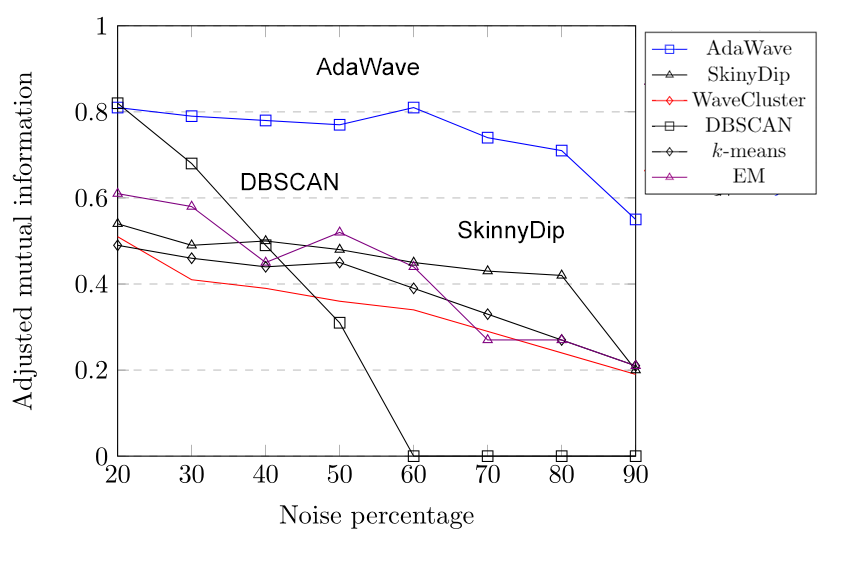}
\vspace{-1em}
\caption{Experimental results on synthetic dataset.}
\label{fig:cluster_result}
	\vspace{-0.5em}	
\end{figure}

Fig. \ref{fig:cluster_result} presents the results of AdaWave and the baselines on the synthetic data. 
%A point corresponds to the mean AMI value from all datasets corresponding to that parameter combination. 
With regard to fairness of the techniques that have no aware of noise (e.g. centroid-based techniques), the AMI only considers the objects which truly belong to a cluster (non-noise points). %The results are discussed in Section \uppercase\expandafter{\romannumeral6}. 
%, and the error bars span one standard deviation in each direction (omitted on every second point to reduce clutter). 
AdaWave clearly outperforms all the baselines for every parameter setting of the noise, and is less sensitive to the noise increase. 
Even with 90\% noise, AdaWave still has a high AMI of 0.55 while others are around 0.20 except DBSCAN. 
EM, SkinnyDip, $k$-means and WaveCluster behave similarly, while WaveCluster yields the lowest, and $k$-means the second lowest. 
DBSCAN can be as good as AdaWave in low noise environment (20\%), but its performance decays quickly and it could not find any clusters when the noise is above 60\%. 

	\vspace{-0.5em}	

\subsection{Real-World Datasets}
For real-world data, we analyzed nine datasets of varying size from the UCI\footnote{http://archive.ics.uci.edu/ml/} repository, namely Seeds, Roadmap, Iris, Glass, DUMDH, HTRU2, Dermatology (Derm.), Motor and Wholesale customers (Whol.). These classification-style data are often used to quantitatively evaluate clustering techniques in real-world settings and the nine datasets we use include a 2D map data and higher dimensional datasets. For some real data where every point is assigned a semantic class label (none of the data include a noise label), we run the $k$-means iteration (based on Euclidean distance) on the final AdaWave result to assign every detected noise objects to a ``true'' cluster. Class information is used as ground truth for the validation.

Table \uppercase\expandafter{\romannumeral1} summarizes the results for different datasets ($n$ denotes the number of data points and $d$ the dimension of the data). In a quantitative sense, AdaWave's results are promising when comparing to the baselines. It achieves the highest AMI value on six of the nine datasets, and ranks the third on two of the remaining datasets (Iris and Whol.). Only on the Seeds dataset AdaWave ranks the fourth among the eight algorithms.
In general, AdaWave behaves the best with an average AMI of 0.60, followed by $k$-means, SkinnyDip and STSC with an average AMI of around 0.49. 

\begin{table*}[h]
\caption{Experimental result on real-world datasets.}
\centerline{
\begin{tabular}{|c|c|c|c|c|c|c|c|c|c|c|}
  \hline
			\makecell*[tr]{Dataset\\($n$,$d$)}&\makecell*[tr]{Seeds\\210,7}&\makecell*[tr]{Roadmap\\434874,2}&\makecell*[tr]{Iris\\150,4}&\makecell*[tr]{Glass\\214,9}&\makecell*[tr]{DUMDH\\869,13}&\makecell*[tr]{HTRU2\\17898,9}&\makecell*[tr]{Derm.\\366,33}&\makecell*[tr]{Motor\\94,3}&\makecell*[tr]{Whol.\\440,8}& \makecell*[tr]{AVG \\ }\\
			\hline
           \makecell[tr] {AdaWave}    &0.475           &\textbf{0.735} & 0.663        &\textbf{0.467}&\textbf{0.470}&\textbf{0.217}&\textbf{0.667}  &\textbf{1.000}  &0.735    & \textbf{0.603} \\
            \makecell[tr]{SkinnyDip}  & 0.348               & 0.484    & 0.306        &	0.268           &0.348    &0.154   &0.638                   &\textbf{1.000}  &\textbf{0.866}	&0.490\\
            \makecell[tr]{DBSCAN}     & 0.000	            & 0.313    & 0.604	      &0.170            &0.073    &0.000   &0.620                   &\textbf{1.000}  &0.696	&0.386\\
            \makecell[tr]{EM}         & 0.512               & 0.246    &\textbf{0.750}&0.243            &0.343    &0.151   &0.336                   &0.705            &0.578	&0.429\\
           \makecell[tr] {$k$-means}	  & \textbf{0.607}      & 0.619	   & 0.601	      &0.136      		&0.213    &0.116   &0.465                   &0.835            &0.826	&0.491\\
           \makecell[tr] {STSC}	      & 0.523               & 0.564	   &0.734         &0.367	        &*0.000   &*0.000  &0.608                   &\textbf{1.000}   &0.568	&0.485\\
            \makecell[tr]{DipMean}	  & 0.000	            &0.459     &0.657	      &0.135		    &0.000    &*0.000  &0.296                   &\textbf{1.000}   &0.426	&0.330\\
            \makecell[tr]{RIC}	      & 0.003	            &0.001     &0.424	      &0.350		    &0.131    &0.000   &0.053                   &0.522            &0.308	&0.199	\\
            \hline
            \multicolumn{10}{l}{$^{\mathrm{*}}$The three settings with * failed due to non-trivial bugs in provided implementations.}
	\vspace{-0.5em}	            
\end{tabular}}

%\caption{Experimental result}
		\label{tab:Margin_settings}
       \end{table*}

\subsection{Case Study}
As a qualitative case study, we investigate two of the clustering results in detail. 

\begin{figure}[ht]
\centering
\includegraphics[scale=0.35]{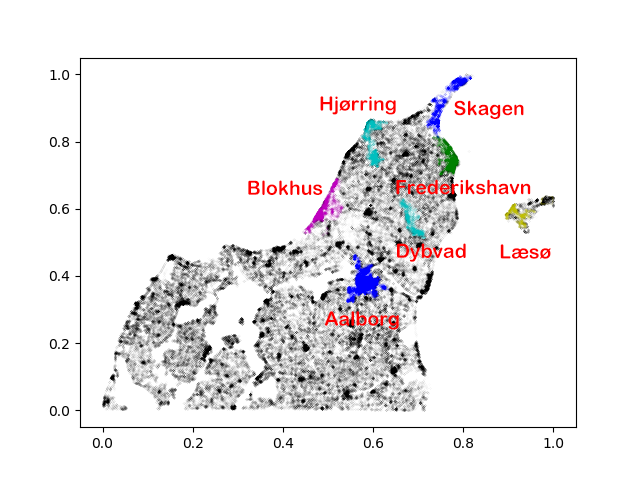}
\vspace{-1em}
\caption{AdaWave result on Roadmap.}
\label{fig:AdaWaveOnRoadmap}
\end{figure}

Roadmap dataset was constructed by adding elevation information to a 2D road network in North Jutland, Denmark (covering a region of 185 x 135 $km^2$). In this experiment, we choose the original 2D road network as the dataset for clustering. The horizontal and vertical axes in Fig. \ref{fig:AdaWaveOnRoadmap} represent latitude and longitude respectively and every data point represents a road segment. Roadmap is clearly a typical highly noisy dataset because the majority of road segments can be termed as ``noise'': long arterials connecting cities, or less-dense road networks in the relatively sparse-populated countryside. In our AdaWave algorithm, we apply 2D $DWT$ on the highly noisy Roadmap and further filters on the transformed grids, so that dense groups are automatically detected (with a highest AMI value of 0.735). The clusters AdaWave detected are generally highly-populated areas (cities like Aalborg, Hj\o rring and Frederikshavn with populations over 20,000), which also verify the correctness of our result.

\vspace{-0.5em}	

The second example is the Glass identification. There are nine attributes: refractive index and 8 chemical elements (Na, Mg etc.). As the dimension is relatively high and the correlation of some attribute to the class is weak (shown in Table \uppercase\expandafter{\romannumeral2}), most techniques produce poor performances for the Glass classification (the AMI value is less than 0.3). Instead of projected in all directions independently, AdaWave detects the connected 3 grids in the 9 dimensional feature space after the discrete wavelet transform, where a one-dimensional wavelet transform will be applied nine times. Though the clustering result of AdaWave with an AMI value of 0.467 is not particularly good, it is considerably much better than the baselines. The results of Glass also reveals the difficulty of clustering in high-dimension data that has weakly correlation with class in each separate dimension.

\begin{table}[htbp]
\caption{Each attribute's correlation with class (Glass)}
\begin{center}
\begin{tabular}{|c|c|c|c|c|c|}
\hline
  \textbf{Attribute} &RI &Na &Mg &Al &Si\\
\hline 
  \textbf{Corelation} &-0.1642 &0.5030 &-0.7447 &0.5988 & 0.1515\\
\hline
  \textbf{Attribute} &K &Ca &Ba &Fe&\\
\hline 
  \textbf{Corelation} &-0.0100 &0.0007 &0.5751 &-0.1879&\\
\hline
\end{tabular}
\label{tab1}
\vspace{-0.5em}	
\end{center}
\end{table}

\subsection{Runtime comparison}
To further explore the efficiency of AdaWave, we carry out runtime experiment on synthetic data with scale $n$ (the number of objects) and compare with several typical clustering algorithms. AdaWave is implemented in python, SkinnyDip is provided in R, and the remaining algorithms are provided in Java. Due to the difference of languages (interpreted vs compiled languages), we focus only on the asymptotic trends.  

\begin{figure}[ht]
\centering
\includegraphics[scale=0.35]{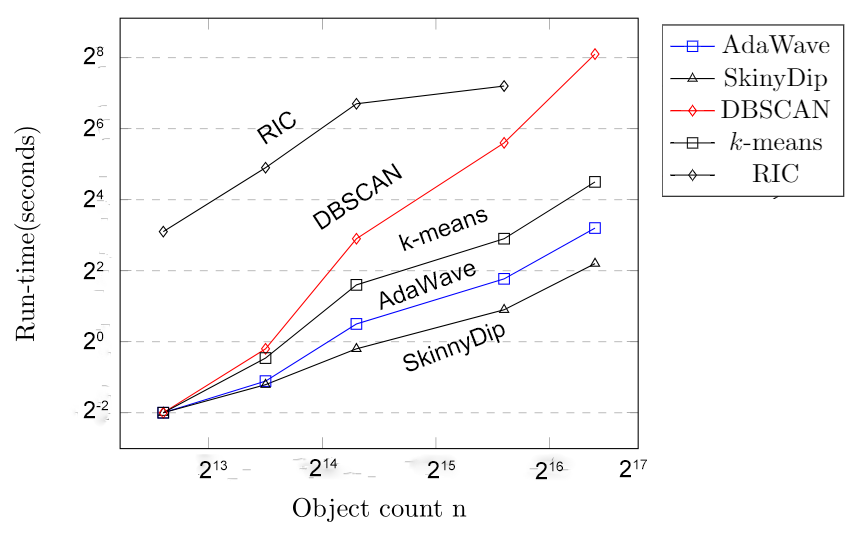}
\vspace{-1em}
\caption{Comparison on running time with several typical algorithms.}
\label{fig:runtime_result}
\end{figure}

We generate the experimental data by increasing the number of objects in each cluster in equal cases. By this means, we are able to scale $n$ (the noise percentage is fixed at 75\%). Then, the experiments are conducted on a computer with Intel Core i5 Processor 4GHz CPU, 500GB Memory, and Windows 10 Professional Service Operation System. 

Fig. \ref{fig:runtime_result} illustrates the comparison on running time, AdaWave ranks the second among the five algorithms with regard to the cost. Comparing to methods like $k$-means and DBSCAN that have a computation complexity of $O(n^2)$, AdaWave is much faster. Although the cost of AdaWave is slightly higher than SkinnyDip, which grows sub-linearly, AdaWave gains a much higher AMI in such situation. In other words, the high efficiency of SkinnyDip clustering is at the cost of shape limitation.

%According to our experimental result (Fig.\ref{fig:runtime_result}), though the cost of AdaWave is slightly inferior to SkinnyDip, which grows sub-linearly, SkinnyDip cannot gain high AMI in such situation. In other words, the high efficiency of SkinnyDip clustering is at the cost of shape limitation. Compared to the other methods like k-means and DBSCAN, which have the computation complexity of $O(n^2)$, AdaWave shows its strength in run time, ranking the second among the five algorithms.

In general, AdaWave performs well in practical run-time growth. AdaWave is essentially a grid-based algorithm. The number of grids is much less than that of the objects, especially when the dimension is high. Due to process of quantization and the linear complexity of wavelet, AdaWave can eventually outperform other algorithms in runtime experiments.

\section{Discussion}
\label{sec:Discuss}
In this section, we discuss in detail on AdaWave and the baselines to have a deeper understanding on these clustering algorithms. %As in the previous section, the baselines include representatives of clustering families and some state-of-the-art automatic clustering methods, which enable us to discuss AdaWave in depth.
We first make observation on the performances of the baselines. 

\begin{itemize} 
	\setlength{\itemsep}{0pt}
	\setlength{\parsep}{0pt}
	\setlength{\parskip}{0pt}	
	\item The standard $k$-means~\cite{Steinhaus1957kmeans,Forgy1965kmeans} tends to lack a clear notion of ``noise'' and has a low AMI value when applied to synthetic data even when the correct $k$ is given. For real data, as the noise is not very high, $k$-means yields the second best on average. 
	\item Model-based approaches like EM~\cite{CeleuxG1992Elsevier} also fail to gain good performance when the cluster shapes do not fit a simple model. For Roadmap dataset which is of irregular shape, EM performs poorly and is the second worst of the 8 methods.
	\item The popular DBSCAN~\cite{DBSCAN1996KDD} method is known to robustly discover clusters of varying shapes in noisy environment. With the increase of noise, however, the limitation of DBSCAN (finding clusters of varying density) is magnified when the dataset also contains large amount of noise. As shown in Fig. \ref{fig:cluster_result}, the AMI of DBSCAN suffers a sharp decline when the noise percentage is above 20\% and DBSCAN performs much worse than the others in extremely noisy environments. 
	\item STSC~\cite{bohm2010clustering} and DipMeans~\cite{kalogeratos2012dip} are two well-known approaches of automatic clustering, but they failed due to some nontrivial bugs or get an AMI of zero for some real datasets. They show similar performance as $k$-means for the synthetic data. (To clarify the figure we did not show their curves).  % but they are unable to differentiate themselves from the other techniques in the experiments (particularly on the challenging, high-noise synthetic cases)
	\item RIC\cite{bohm2006robust} is designed to be a wrapper for arbitrary clustering techniques. Given a preliminary clustering, RIC purifies these clusters from noise and adjusts itself based on information-theoretical concepts. Unfortunately it meets difficulties when the amount of noise in the dataset is non-trivial: for almost all of our experiments with noisy data, the number of clusters detected is one with a corresponding AMI value of zero. 
	\item The dip-based technique SkinnyDip~\cite{maurus2016skinny}, which is specialized to deal with high noise data, also meets challenge in the synthetic experiment because the projection of clusters in every dimension is usually not a unimodal shape as the algorithm desires.  In general, SkinnyDip is better than all other baselines on both synthetic and real data. However, due to its strict precondition, SkinnyDip does not work as well as AdaWave in real-world data.
\end{itemize}

%selecting the appropriate scale of analysis and the number of clusters using a local scale approach with affinities. STSC obtained a good result on the Pen-Digits data set,  

%DipMeans, it is also clear that such techniques do not have a clear concept of noise, but even if we ignore noise in our quality measurements (as in our synthetic experiments) we still see these techniques fail because of model- and distance-assumptions.

In general, mainly due to the process of wavelet transform and threshold filtering, AdaWave outperforms the baselines on synthetic and real-world data by a large margin. The margin is more or less maintained as the number of clusters and level of noise vary. 
The wavelet filters emphasize regions where points are densely located and simultaneously suppress weaker information on the boundary, thereby automatically removing the outliers and making the clusters more salient. Threshold filtering further uncovers true clusters from dense-noise grids to ensure the strong robustness to noise. The main reason for the inferiority of the baselines can be seen via an investigation on the ring-shape case: the clusters are in two overlapping circular distributions with dense noise around, for which the comparison methods tend to group together as one or separate them as rectangle-style clusters (causing a zero AMI). Applying DWT to wavelet feature space to process coefficient reduction and threshold filtering proves to be a feasible solution for such problems.

Like other grid-based techniques, with the increase on the dimension $d$, the limitations of AdaWave begin to emerge. Due to the high dimension, the number of grid cells rises sharply (exponential increase) when rasterized in every dimension. Grid-based clustering tends to be ineffective due to the high computational complexity. However, our new data structure provides us a storage-friendly solution that can handle the high dimensional data which is sparse in most real applications. 
%; however, it cannot avoid high-dimension problem in the fundamental sense. 
%On the other hand, by our experimental results, AdaWave yields low performance when the noise percentage is low (less than 10\%) and even inferior to those with nearly 30\% noise. The main reason is that the ``elbow theory'' may not work well in the low-noise situation. 
%Threshold filtering aims at further eliminating dense-noise grids. However, most outliers ($noise$ $grids$) are removed in the process of coefficient reduction, thus influencing the determination of threshold. Practically this rarely happens, because AdaWave focuses on clustering of the datasets with extremely high noise and of arbitrary shape.

\section{Conclusion}

In this paper, we propose a novel clustering algorithm AdaWave for clustering data in extremely noisy environments. AdaWave is a grid-based clustering algorithm implemented by applying a wavelet transform to the quantized feature space. 
AdaWave exhibits strong robustness to noise, outperforming other clustering algorithms including SkinnyDip, DBSCAN and $k$-means on both synthetic data and natural data. Also, AdaWave doesn't require the computation of pair-wise distance, resulting in a \textit{low complexity}. On the other hand, by deriving the ``grid labeling'' method, we drastically reduce the memory consumption of wavelet transform and thus AdaWave can be used in analyzing dataset in relatively \textit{high dimension}. Furthermore, AdaWave is able to detect \textit{arbitrary clusters}, which the SkinnyDip algorithm cannot. By wavelet transform, AdaWave inherits the ability to analyze data in \textit{different resolutions}.
 Such properties enable AdaWave to fundamentally distinguish itself from the centroid-, distribution-, density- and connectivity-based approaches. 
 %AdaWave outperforms competing techniques especially when clusters are of irregular shapes and the environment is extremely noisy. 

%\section*{Acknowledgment}
%This work is supported in part by NSFC 61772219, and MSRA Collaborative Research Program.

%\vspace{12pt}
%\color{red}
%IEEE conference templates contain guidance text for composing and formatting conference papers. Please ensure that all template text is removed from your conference paper prior to submission to the conference. Failure to remove the template text from your paper may result in your paper not being published.

\end{document}